\documentclass{aa}  

\def\kpc{{\ \rm kpc}\ h^{-1} }
\def\kms{{\ \rm km \ s^{-1}}}

\usepackage{graphicx}
\usepackage{caption}
\usepackage{subcaption}
\usepackage{txfonts}
\usepackage[dvipsnames]{xcolor} 
\begin{document}

   \title{Galaxy Pairs in Cosmic Voids}

   \author{Maria Laura Ceccarelli\inst{1,3},
     Sol Alonso\inst{2},
        \and 
         Diego Garcia Lambas\inst{1,3}
                    }

   \institute{
    Instituto de Astronom\'{\i}a Te\'orica y Experimental, (IATE-CONICET),
Laprida 854, X5000BGR, C\'ordoba, Argentina\\
              \email{laura.ceccarelli@unc.edu.ar},
            \and
            Departamento de Geof\'{i}sica y Astronom\'{i}a, CONICET, Facultad de Ciencias Exactas, F\'{i}sicas y Naturales, Universidad Nacional de San Juan, Av. Ignacio de la Roza 590 (O), J5402DCS, Rivadavia, San Juan, Argentina            
\and
         Observatorio Astron\'omico de C\'ordoba, Universidad Nacional de C\'ordoba, Laprida 854, X5000BGR, C\'ordoba, Argentina\\
             }

   \date{Received xxx; accepted xxx}

 

   { }

   \abstract{
We present a statistical analysis of different astrophysical properties of a sample of galaxy pairs in cosmic voids. The sample consists of 72 galaxy pairs with projected separations and relative radial velocities rp<100 h$^{-1}$kpc, $\Delta V <$ 500 $\kms$ in the redshift range z<0.1. The different results for this pair sample are compared to those derived for matched samples configured in absolute magnitude, stellar mass and concentration residing in void wall and global averaged environments.
We find that pair galaxies in voids tend to have bluer optical colors than the corresponding galaxies in wall an field, regardless of their stellar mass and concentration, which indicates a more recent formation of the bulk of stars.
We also obtain larger mid--IR colors for the void paired galaxies with respect to the corresponding matched samples in the wall and in field environments. However, we find significantly larger differences for galaxies with high mass and concentration.
We also notice that mid--IR color--color diagram shows void pair members consistent with the locus of star--forming galaxies, in contrast with the other environments that exhibit a bimodal behavior comprising both  passive and star--forming objects.
The D$_n$(4000) parameter also  shows a significant younger stellar population in paired galaxies in voids. This is also reflected in the higher star formation rate values, which show a larger efficiency for void paired galaxies.
We notice that the star formation efficiency is larger for void paired galaxies with high stellar mass and concentration. 
We also find that the efficiency of star formation associated to galaxy interactions is significantly larger in pairs residing in cosmic voids. This larger star formation activity could be associated to both the expected richer gas environment and a more gentle dynamical behavior as compared to the more eccentric orbits and stronger interactions and mergers more likely to occur in wall and field environments. 
   }
   
   \keywords{galaxy interactions --
                cosmic voids --
                SDSS
               }

   \maketitle

%

\section{Introduction}
    
    \vspace{0.5cm}
Galactic interactions serve as a powerful mechanism for altering different galaxy properties, primarily triggering star formation activity \citep{alo06,alo12,Barton2000,Lambas2003,Lambas2012,Kenni98,Mesa2014,pear19,pear22} and reshaping both galaxy morphology and the stellar mass function.
The underlying physical mechanisms of galaxy-galaxy interactions have been elucidated through theoretical and numerical analyses 
\citep{Mart95,Toom72,Bar92,Bar96,Mih96}. These studies reveal that tidal torques generated during close encounters lead to collisional disruption, material dissipation, and gas inflows.
Moreover, galaxy mergers contribute to the growth of supermassive black holes at galactic centers \citep{alonso2007, dotti12,elli19}, with the efficacy of this process contingent upon the gas reservoir and internal characteristics of the interacting galaxies.
Collectively, these effects underscore the significance of galaxy mergers in the evolutionary trajectory of galaxies.
In this context, according to hierarchical structure formation models,  galaxy mergers and interactions play a crucial role in the structure formation scenario of the Universe.

Moreover, these effects are heavily contingent upon the specific local environments where galaxy interactions take place \citep{alo06,alo12,Lambas2003,Lambas2012,das21}, with a preference for occurrences within group environments \citep{barton2007,McI08}.
While a robust correlation between the rate of galaxy mergers and the environment was observed at high redshifts (z $\approx$ 1), with elevated merger rates in dense regions \citep{lin10,deRa11}, this dependence weakened considerably in the local Universe \citep{darg10,elli10}. Galaxies within massive cluster environments exhibit fewer signs of interactions and mergers compared to those in groups or fields 
\citep{McI08,tran08,alo12,kam13}, primarily due to the high relative velocities of galaxies in clusters, inhibiting mergers \citep{deger18,bena20,sure24}. In addition, \cite{skib09} observed that the likelihood of a galaxy undergoing a merger or interaction is largely uncorrelated with the environment, except at projected separation scales of 100 $\kpc$. Intermediate density environments, as highlighted by \cite{Perez2009}, foster efficient close encounters and mergers. 
More recently, \cite{sure24} found that mergers exhibit a preference to occur in the underdense regions on scales greater than 50 $\kpc$ of the large scale structure.
Simulation studies also support an environmental dependence of mergers, with semi-analytic models showing varying strengths of this dependence \citep{jian12}. Additionally, simulations suggest that galaxy pairs tend to avoid extreme environments such as clusters and cosmic voids \citep{tonn12}.

Extremely low-density regions termed cosmic voids emerge as prominent 
components of the large--scale Universe.  
They are an integral part of a more extensive and complex network, namely the cosmic web.
Although voids make up more than 40\% of the total volume in galaxy surveys, their low galaxy density (lower than 20\% the average in density the Universe) limits the availability of suitable samples for statistical analysis \citep{pan2012, cecca13} . Therefore, studying voids requires large surveys covering vast areas and deep depths.
The study of galaxies in voids have been carried out, the results of which show objects in primitive stages, consistent with a slower evolution of galaxies in voids \citep{beygu2016, beygu2017}. 
Mostly devoid of galaxies, the few structures that inhabit them are characterized by higher star--formation activity, bluer colors, lower mass, and late type morphology, with significant amounts of gas,  
when compared to galaxies in an average density environment 
\citep{grogin2000,
parkprop2007,von_benda-beckmann_void_2008,kreckel2011,ricciardelli2014,liu2015,moorman2016,beygu2016,beygu2017,ricciardelli2017}. 

Moreover, cosmic voids are exceptional regions filled with pristine gas, with unique dynamic conditions characterized by velocity fields dominated by divergent flows with low velocity dispersion \citep{padilla2005,ceccarelli2006,paz_clues_2013,hamaus2015, correa2022}. 
Due to the enormous extension of these structures, it is reasonable to expect that the galaxies that inhabit them have formed and evolved in this environment. 
In this context, voids are positioned as scenarios with formidable potential for understanding the formation and evolution of galaxies, including the effects of interactions.
 
The particular environment and dynamics of cosmic void interiors has important consequences for the evolution of galaxy pairs.  In effect, given a present-day pair of interacting galaxies in a regular environment, their member galaxies have 
probably been separated by several Mpc at early times and have a large negative pairwise velocity. 
On the contrary, pairs in a cosmic void have a significantly lower pairwise velocities with the global dynamics of the pair as a whole, dominated by the void expansion giving its imprint on galaxy orbits. 
Thus, it is expected that the evolution of void pair member astrophysical properties
may be significantly different and deserve an appropriate analysis.  
In this direction, acquiring and analyzing paired galaxies situated within cosmic voids poses a significant challenge due to their low galactic density and specific dynamics. 
Nevertheless, these particular environments offer an exceptional opportunity to explore the study of paired systems and galactic interactions, devoid of external influences. 
This will provides an invaluable chance to gain profound insights into the impact of interactions on galactic properties and their evolutionary processes.

The paper follows this structure: Section 2 outlines the methodology for selecting pair systems within cosmic voids. 
In Section 3 we obtain carefully pair galaxies in wall and field environments exhibited comparable redshift, luminosity, mass and morphology as the galaxy pairs located in void regions.
Section 4 presents the analysis of void environment effects on paired galaxies, with a detailed examination of their colors, star formation rates, and stellar populations in comparison with pair systems located in wall and field.
Section 5 discusses the findings and provides a summary of the key conclusions.

Throughout this paper we adopt a cosmological model characterized by the parameters $\Omega_m=0.3$, $\Omega_{\Lambda}=0.7$ and $H_0=70 \kms \rm Mpc ^{-1}$.

\section{Selection of pair systems in cosmic voids}

The data utilized in this study has been sourced from the Sloan Digital Sky Survey (SDSS), one of the most remarkable astronomical survey \citep[SDSS;][]{York2000}. Spanning several operational phases (SDSS-I, 2000-2005; SDSS-II, 2005-2008; SDSS-III, 2008-2014), the SDSS has consistently released its data to the scientific community on an annual basis. With the advent of the latest iteration, SDSS-IV (2014-2020) \citep[SDSS-IV;][]{sdssiv}, the survey has entered a new realm, pushing the boundaries of precision cosmological measurements into the pivotal early epochs of cosmic evolution through eBOSS. Additionally, it has expanded its infrared spectroscopic survey of the Galaxy across both northern and southern hemispheres with APOGEE-2, and has introduced the Sloan spectrograph to create spatially resolved maps of individual galaxies through MaNGA.

For this work we employ a galaxy pair sample obtained from \cite{mesa18}. 
To identify these pairs, they selected galaxies with a projected separation of $r_{p}<$ 100 $\kpc$ and relative radial velocities $\Delta V <$ 500 $\kms$, limited to $z<0.1$. 
These criteria yielded an sample of 25965 galaxy pairs.
The data used to compile this galaxy pair sample were sourced from the Main Galaxy Sample \citep[MGS;][]{strauss2002}, obtained from the \texttt{fits} files on the SDSS website\footnote{http://www.sdss3.org/dr8/spectro/spectro$\_$access.php}. 
For this sample, k-corrections were performed and band-shifted to $z=0.1$ using the \texttt{k-correct\_v4.2} software by \cite{Blanton2007}. K-corrected absolute magnitudes were derived from Petrosian apparent magnitudes, converted to the AB system.
 
In addition, for this study we use a catalogue of 194 voids of \cite{cecca13} identified in the Main Galaxy Sample of the SDSS data release 7 (DR7). In order to avoid luminosity biases on voids, they are identified in a volume complete sample selected accordingly the catalogue limiting magnitude. This catalogue features the right ascension, declination, and redshift of the center of each void as well as the radius in h$^{-1}$ Mpc ($R_{v}$). The radius ranges from 5 to 22 h$^{-1}$ Mpc and the the maximum redshift is $z = 0.12$.   
 
In order to locate galaxy pairs within cosmic voids, we conducted a cross-correlation between the sample of paired galaxies and the void catalog.
With this aim we compute the cosmological distances across between the paired galaxies and the center of each void, requiring that said distance be less than the size of the void.
We have also defined the following regions: 

- void interior $\it(void)$: from 0 to 0.8 $R_{v}$ 

- void wall $\it(wall)$: from 0.8 to 1.2 $R_{v}$

Moreover, aiming to contrast the impact of galactic interactions within cosmic voids against those in typical environments, we selected paired galaxies in the field, considering:

- field environment $\it(field)$: from 1.8 to 3.0 $R_{v}$ 

The first two zones are related to the cosmic voids and an additional region is situated in the field outside the voids.
In addition, the void interior and wall, comprises underdense regions whereas the field correspond to average global densities and is used for comparison purposes.

Following this procedure we found 72, 790 and 10411 galaxy pairs in the center, wall and field respectively (see Table \ref{tab:Table1}).

\begin{table}
\center
\caption{Restrictions and numbers of total galaxy pairs and matched pairs in the different regions.}
\begin{tabular}{|c c c c| }
\hline
Regions & Restrictions &  Total Pairs & Matched Pairs   \\
\hline
\hline
\it{void interior} & 0 to 0.8 $R_{v}$ & 72 & 72 \\
\it{void wall} &  0.8 to 1.2 $R_{v}$ & 790  & 226 \\
\it{field} & 1.8 to 3.0 $R_{v}$ & 13411 & 4395 \\
\hline
\hline
\end{tabular}
{\small}
\label{tab:Table1}
\end{table}

\section{Matched Pair Samples}

\begin{figure}[ht]
\begin{center}
\includegraphics[width=0.5\textwidth]{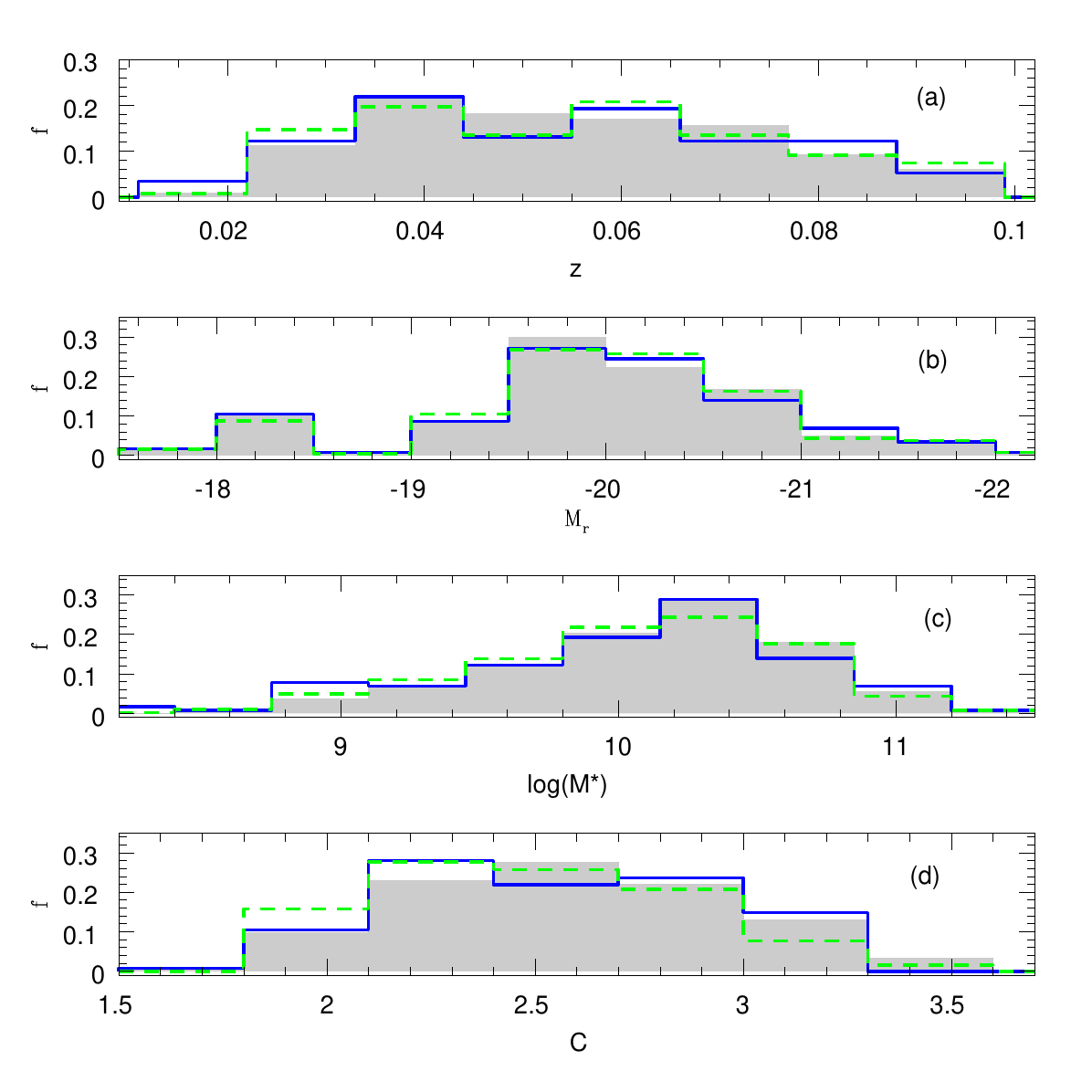}
\caption{Normalized distributions of redshift, ${r-}$ band absolute magnitude, stellar masses and concentration parameter for pair galaxies in voids (blue solid lines), in wall (green dashed lines) and in fields (shaded histograms).}
\label{control}
\end{center}
\end{figure}

With the aim to focus our attention on the impact of the particular void environment on galaxy interactions,  we employed a Monte Carlo algorithm to obtain carefully pair galaxies in wall and field environments exhibited comparable distributions of redshift, r$-$band absolute magnitude, and stellar masses as the void galaxy pairs in our sample
(see panels a, b and c in Fig.~\ref{control}). 

Moreover, taking these considerations into account, we found that paired galaxies located in walls and fields equal their concentration indices, C\footnote{$C=r90/r50$ is the ratio of Petrosian 90 \%- 50\% r-band light radii}, to those of void pairs, as depicted in panel d of Fig.~\ref{control}. 
This parameter discriminates between bulge and disk galaxy types and is also a well tested morphological classification index  \citep{Strateva2001}. Moreover  \citet{Yamauchi2005} conducted a galaxy morphological classification based on the C parameter, revealing a high level of agreement with visual classification methods.
Then, this observation indicates a comparable bulge-disk relationship between the different samples. Thus, any disparities observed in the outcomes can be attributed to environmental influences rather than variations in galaxy morphology.

We also checked that the matched samples and void pair galaxy sample were drawn from the same distributions by applying the Kolmogorov-Smirnov (KS) test and obtaining p $>$ 0.05 for the null hypothesis.
The methodology employed to obtain matched samples of paired galaxies in wall and field environments ensures that it shares the same selection effects as the void pair catalog. This process facilitates the study of pair features in different environments, elucidating the pure and intrinsic effects of the interaction process that unfold within the cosmic void environment.

Finally, the matched samples of paired galaxies, situated on the walls of the voids and in the field, each comprise 226 and 4395 pairs, respectively (see last column of the Table \ref{tab:Table1}).
These catalogs will be used in the subsequent analyzes outlined in Section 4.

\section{Analysis of void environment impact on paired galaxies}

In this work, our focus lies in exploring the properties of galaxy pair systems situated within cosmic voids, where galactic density is extremely low. 
This particular environment offers an exceptional opportunity to explore the features of pair galaxies, devoid of external influences, 
thus offering deep insights into the effects of interactions on galaxy properties and their evolution.
In this context, the subsequent subsections will delve into the examination of color, age of stellar population and star formation activity of galaxy pairs residing in voids, in comparison with the pair systems inhabiting the void walls and field regions.

\subsection{Effects on optical and infrared colors}

\begin{figure}[ht]
\begin{center}
\includegraphics[width=0.43\textwidth]{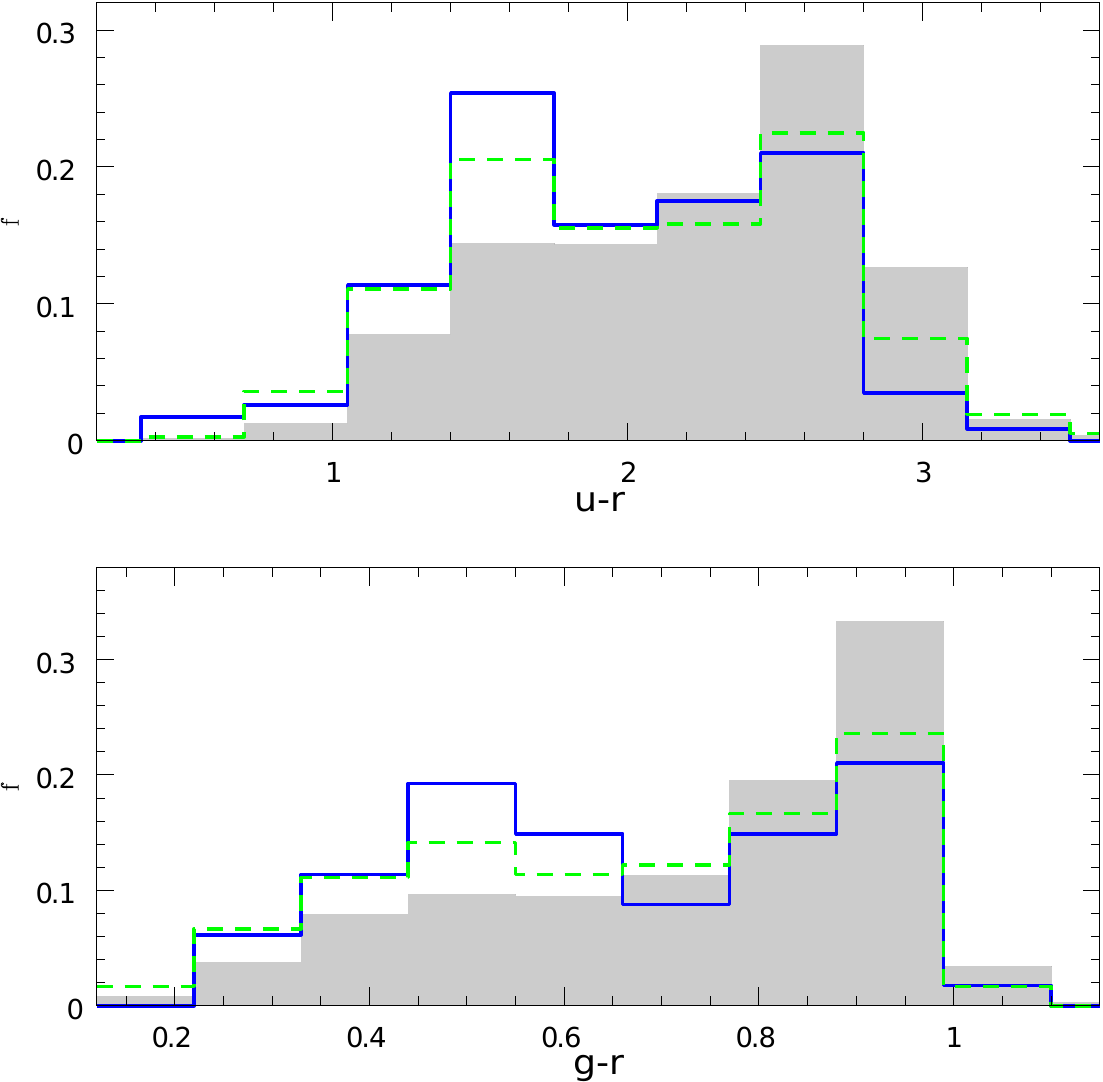}
\caption{Distributions of $u-r$ and $g-r$
for pair galaxies in voids (blue solid lines), in wall (green dashed lines) and in fields (shaded histograms).}
\label{Hcol}
\end{center}
\end{figure}

\begin{figure*}
  \begin{subfigure}[t]{0.49\textwidth}
    \includegraphics[width=\textwidth, height=0.75\textwidth]{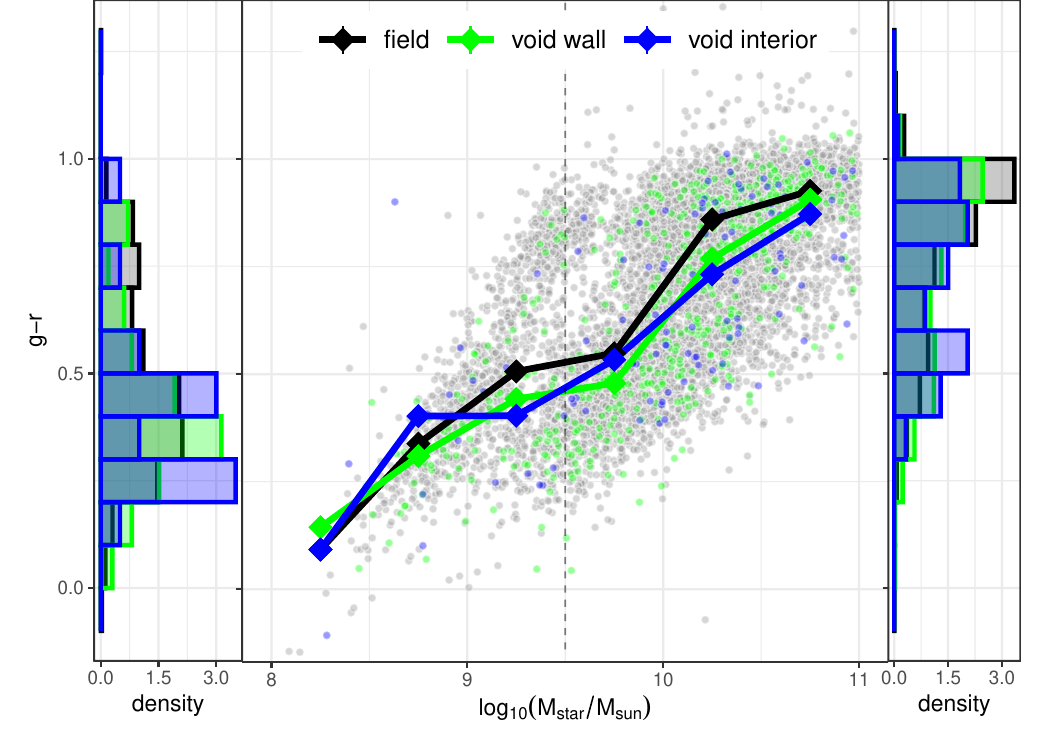}
    \caption{
    {\it{Central panel:}} Scatter plot of $g-r$ as a function of $log(M_*/M_{sun})$ for pair galaxies in 
voids (blue), in wall (green) and in fields (black).
Diamonds and solid lines indicate the median.  
The vertical dotted line represents $log(M_*/M_{sun}) =$ 9.5
{\it{Left panel:}} $g-r$ normalized distribution for $log(M_*/M_{sun}) <$ 9.5 paired galaxies in voids and field. 
{\it{Right panel:}} the same than left panel but for $log(M_*/M_{sun}) >$ 9.5. 
    }
    \label{fig:grvsmstar}
  \end{subfigure}
  \hfill
  \begin{subfigure}[t]{0.49\textwidth}
    \includegraphics[width=\textwidth, height=0.75\textwidth]{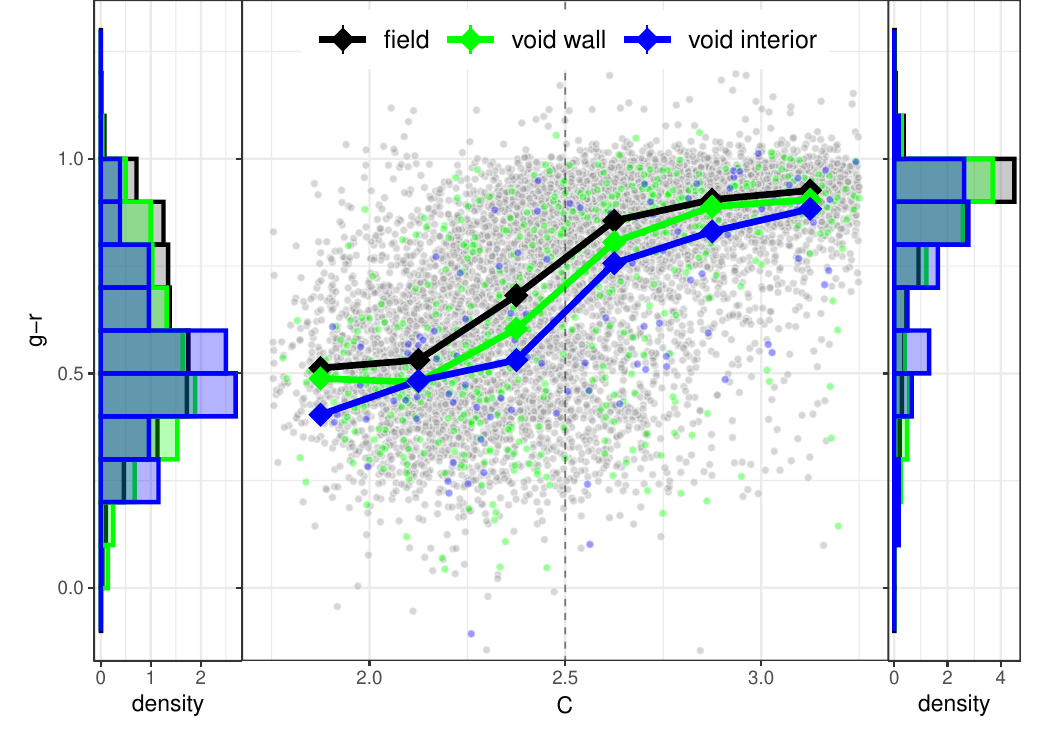}
    \caption{
    {\it{Central panel:}} Scatter plot of $g-r$ as a function of concentration index for 
pair galaxies in voids (blue), in wall (green) and in fields (black).
Diamonds and solid lines indicate the median.
{\it{Left panel:}} $g-r$ normalized distribution for $C <$ 2.5 paired galaxies in voids and field. 
{\it{Right panel:}} the same than left panel but for $C >$ 2.5. 
}
    \label{fig:grvsCI}
  \end{subfigure}
   \caption{Study of stellar mass and concentration impact on color void paired galaxies.  
   }
\end{figure*}

\begin{figure}[ht]
\begin{center}
\includegraphics[width=0.45\textwidth]{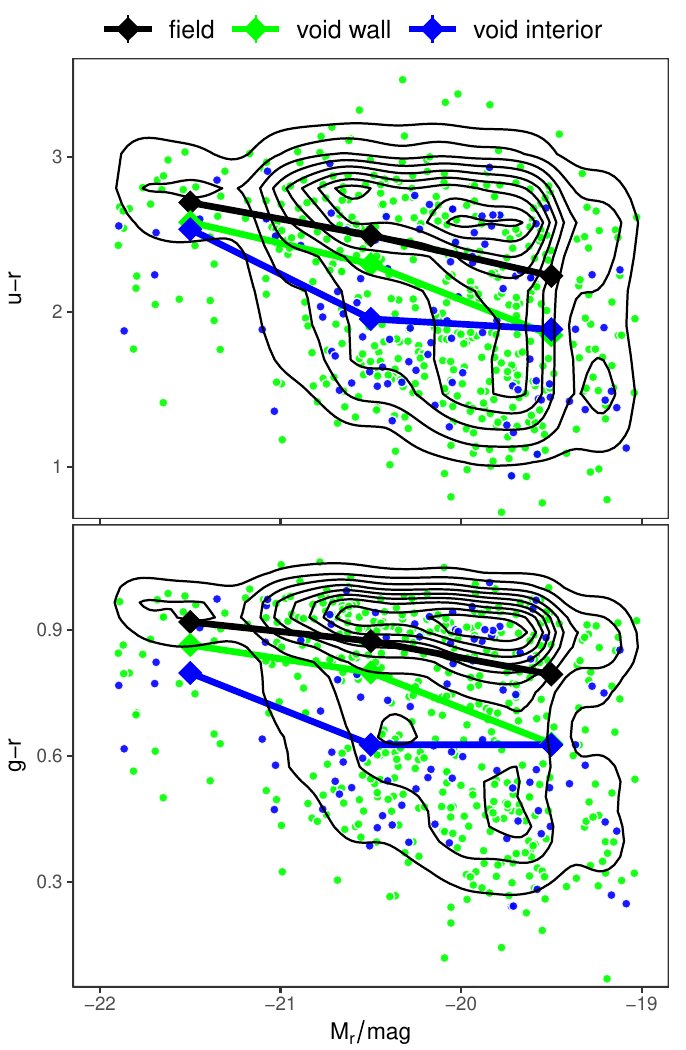}
\caption{Color magnitude diagrams: $u-r$ (upper panel) and $g-r$ (bottom panel) as a function of $M_r$ for pair galaxies located in voids (blue points), in wall (green points) and in fields (black contours). Connected diamonds indicated the median values.
}
\label{ColM}
\end{center}
\end{figure}

Galaxy colors, widely known for their correlation with star formation, stellar population age, and gas content, are also linked to both local and global environments. In this section, our focus is on delving into the colors of paired galaxies situated in cosmic voids. We conduct a detailed analysis of colors within different samples of pairs residing in voids, walls and field environments.

In Fig. \ref{Hcol}, the distributions of both color indices ($\rm u-\rm r$ and $\rm g-\rm r$) are shown for paired galaxies in void, void wall and field environments.  Notably, pair systems situated in cosmic voids show a clear excess of bluer colors with respect to those from the matched samples of pairs residing void walls and fields.
Furthermore, it is observable that the value $ u-r $ $\approx$ 2.2 effectively 
demarcates the two peaks of each color distribution, consistent with findings by \cite{Strateva2001}. This threshold serves to delineate between the blue and red 
galaxy populations. This boundary is also reflected in $ g-r $ $\approx$ 0.7.
To quantify this trend, we examined the surplus of bluer color indices 
($ u-r < $ 2.2 and $ g-r < $ 0.7) among paired galaxies within the 
different environment analyzed in this work, finding a clear excess of bluer colors 
in pair systems located in the inner zones of the cosmic voids 
(refer to Table \ref{tab:Color}).

It is well known that the optical colors of galaxies are linked to their stellar mass 
\citep{cucciati2010}, due to this it is convenient to analyze the relation of color 
indices with stellar mass.
In Fig. \ref{fig:grvsmstar} we show the $g-r$ color as a function of stellar mass, 
for galaxy pairs in void interior (blue circles), void walls (green circles) and in the 
field (grey circles). The lines and diamonds display the corresponding median values 
(field values are shown in black colors). 
We stress that the $C$ and luminosity distributions of the paired galaxies 
in void interior, void wall and field are similar for the different 
stellar mass bins analyzed in this work.
We select high mass ($log(M_*/M_{sun}) > 9.5$) and low mass ($log(M_*/M_{sun}) < 9.5$) 
galaxy samples, 
the vertical dotted line in Fig. \ref{fig:grvsmstar} indicates this separation.
The $g-r$ distributions for these samples according to the large--scale 
environment are shown in the right and left panels, respectively. 
Note that the same color code is used throughout the paper.
As it can be seen in the central panel of the Fig. \ref{fig:grvsmstar}, 
galaxies in void interiors and walls tend to have slightly bluer colors 
than those in the field, 
especially for medium--high stellar masses ($log(M_*/M_{sun}) > 9$).
On the other hand, for low stellar masses ($log(M_*/M_{sun}) < 9$), 
no notable differences are seen in the $g-r$ medians. 
Also note that there is a very low number of galaxies in voids with 
these stellar masses.
Regarding the colors of paired galaxies in the samples of low and 
high stellar--mass, for the former (left panel of Fig. \ref{fig:grvsmstar}) 
the means and medians are comparable in voids and field, 
while for the latter (right panel of Fig. \ref{fig:grvsmstar}) galaxies 
in voids tend to exhibit bluer colors than those in the field.

Furthermore, the color--stellar mass relation depends on the galaxy morphology \citep{sandage1978,Strateva2001,Baldry2004,bundy2010}, 
consequently, studying dependencies on concentration can be useful to analyze the effect 
of paired galaxies on their colors.
In Fig. \ref{fig:grvsCI} we show the $g-r$ as a function of concentration, 
for galaxy pairs in void interior (blue circles), void walls (green circles) and in the 
field (grey circles). The lines and diamonds display the corresponding median values 
(field values are shown in black colors). As it can be noticed in the figure, 
paired galaxies in voids tend to have bluer medians than similar galaxies 
in the field, for all concentrations analyzed.

In Fig. \ref{ColM} the color-magnitude diagrams depicting the relationships between $u-r$ and $g-r$ versus $M_r$ for pair galaxies located in cosmic voids are shown. The corresponding matched samples of pairs found in void walls and fields are also included.
It can be observed that paired galaxies in fields exhibit an excess of them in the \textit{red sequence}.
In contrast, pairs situated in void walls display a distinct bimodality, evident in both the \textit{red sequence} and the \textit{blue cloud}. Furthermore, paired galaxies within void environments tend to cluster towards the \textit{green valley} and the \textit{blue cloud}.

\begin{table*} 
\center
\caption{Percentages of paired galaxies with blue colors located in void interior environment and for matched samples of pairs in void walls and fields. Standard errors are included.
}
\begin{tabular}{|c c c c c| }
\hline
Restrictions &  $ (u-r)<2.2 $ & $ (g-r)<0.7 $ & $(w1-w2)>0.2$ & $(w2-w3)>2.5$\\
\hline
\hline
\% Pairs in Voids &  59.61 $\pm$ 0.72  \% & 53.50 $\pm$ 0.69\% & 33.33 $\pm$ 1.90\% & 70.58 $\pm$ 1.53\% \\
\% Pairs in Walls &  54.42 $\pm$ 0.39 \% & 51.76 $\pm$ 0.38\%  & 26.44 $\pm$ 1.26\% & 66.38 $\pm$ 0.68\% \\
\% Pairs in Fields & 42.64 $\pm$ 0.65 \% & 35.50 $\pm$ 0.59\%  & 20.13 $\pm$ 0.72\% & 52.99 $\pm$ 0.14\% \\
\hline
\hline
\end{tabular}
{\small}
\label{tab:Color}
\end{table*}

Here we explore the mid infrared colors of pair galaxies. The WISE mid IR photometry is appropriate to characterize the emission in these frequencies which have important information on the re emission of light in the mid IR by the dusty environment associated to star forming-regions.

To obtain IR magnitudes we use data from Wide-field Infrared Survey Explorer (WISE,
\citet{wright2010}). We employ the all-sky data release and extract the WISE magnitudes of astronomical sources through matching with our samples of paired galaxies.  
We obtain 102 galaxies in void interior, 470 at void walls and 7673 in the field.
These samples exhibit similar redshift, stellar mass and concentration distributions. 

\begin{figure}[ht]
\begin{center}
\includegraphics[width=0.43\textwidth]{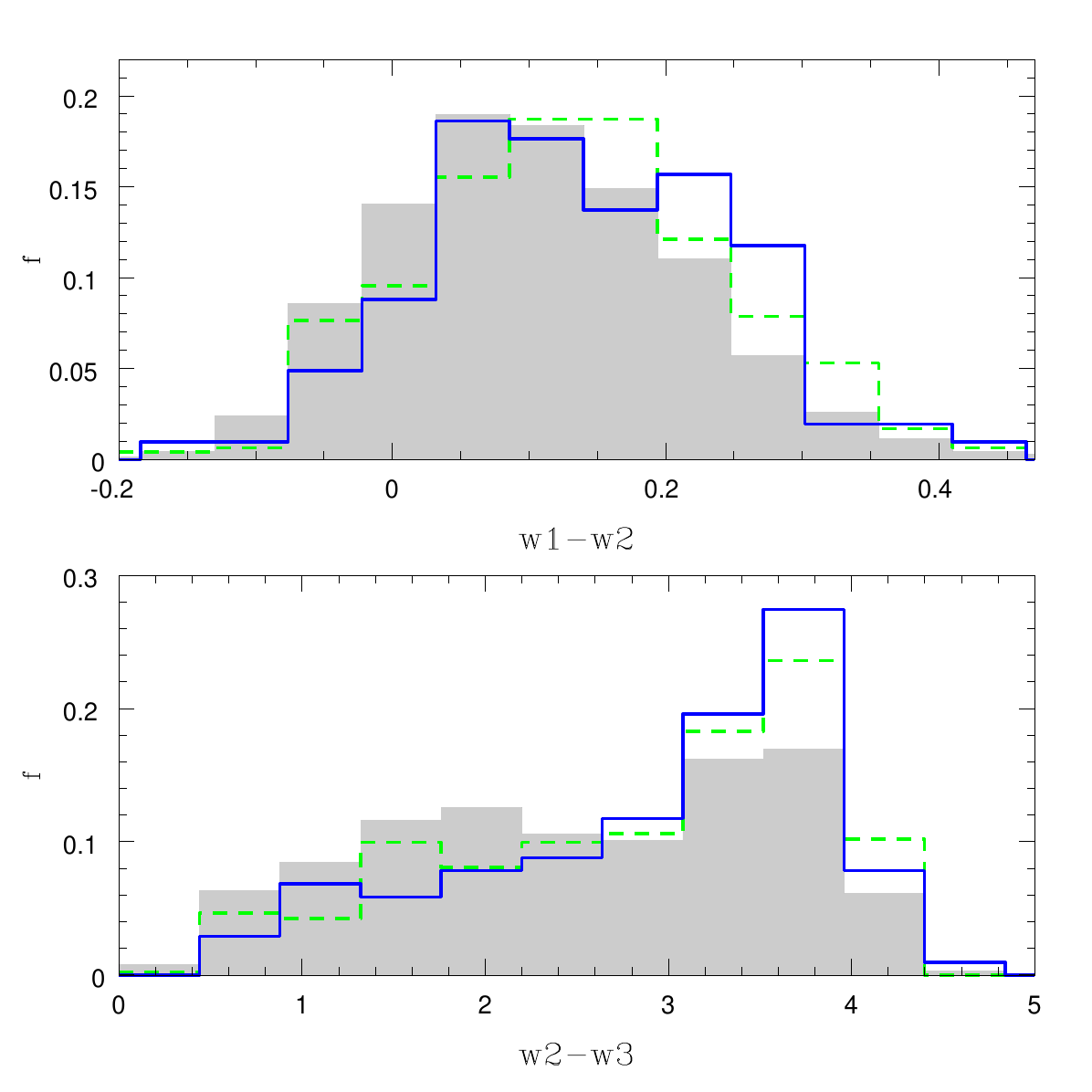}
\caption{Distributions of $w1 - w2$ and $w2 - w3$
for pair galaxies in voids (blue solid lines), in wall (green dashed lines) and in fields (shaded histograms).}
\label{Hcolwise}
\end{center}
\end{figure}

In Fig. \ref{Hcolwise}, the distributions of the infrared color indices ($w1-w2$ and $w2-w3$) are presented for paired galaxies across three environments: voids, void walls, and field. Notably, galaxy pairs located in cosmic voids exhibit higher values for both infrared color indices compared to those in void walls and field environments.
Moreover, to quantify this trend excess, we examined the surplus of infrared color indices ($w1-w2$ > 0.2 and $w2-w3$ > 2.5) among paired galaxies within the different environment studied in this work, finding a clear excess of higher values of infrared colors in pair systems located in the inner zones of the cosmic voids (refer to Table \ref{tab:Color}).
Additionally, we notice in this figure the large values of mid IR colors corresponding to star-forming galaxies, following the works of \citet{wright2010}  and \citet{jarrett17}.

\begin{figure*}[ht]
\begin{center}
\includegraphics[width=0.9\textwidth]{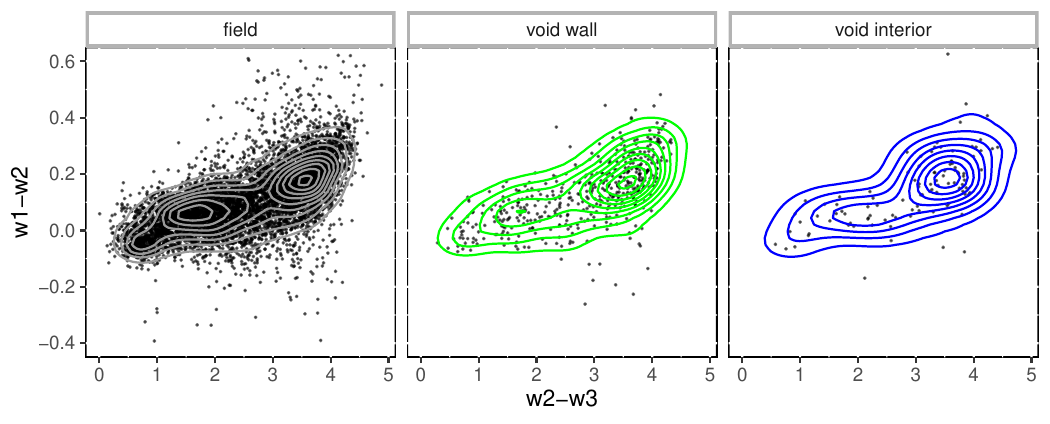}
\caption{WISE color--color diagrams for pair galaxies located in voids (right panel, blue contours), in wall (central panel, green contours) and in fields (left panel, grey).
}
\label{wisecc}
\end{center}
\end{figure*}

We explore the behavior of wise color--color diagram for galaxies in voids, walls and field and we display them in Fig. \ref{wisecc}. Each panel in the figure correspond to a large--scale region as is indicated, black point correspond galaxies and the contours indicate iso--density levels.
The color--color diagram for galaxies inside voids (right) exhibits a single notable maximum for high values of $w1 - w2$ and $w2 - w3$, consistent with a population dominated by galaxies with intense star formation activity. On the other hand, the diagram corresponding to paired galaxies in the field (left panel) presents two prominent maxima, indicating a bimodal behavior, corresponding to passive and star-forming galaxies.
Specifically, the maximum corresponding to lower color values indicates the existence of a population of passive galaxies, while the one exhibiting higher color values indicates a population of galaxies with active star formation.
Regarding paired galaxies in walls (middle panel), they are in an intermediate stage between the field and the interiors of voids, presenting a dominant star forming population and an incipient passive population.

\begin{figure*}
  \begin{subfigure}[t]{0.49\textwidth}
    \includegraphics[width=\textwidth, height=0.75\textwidth]{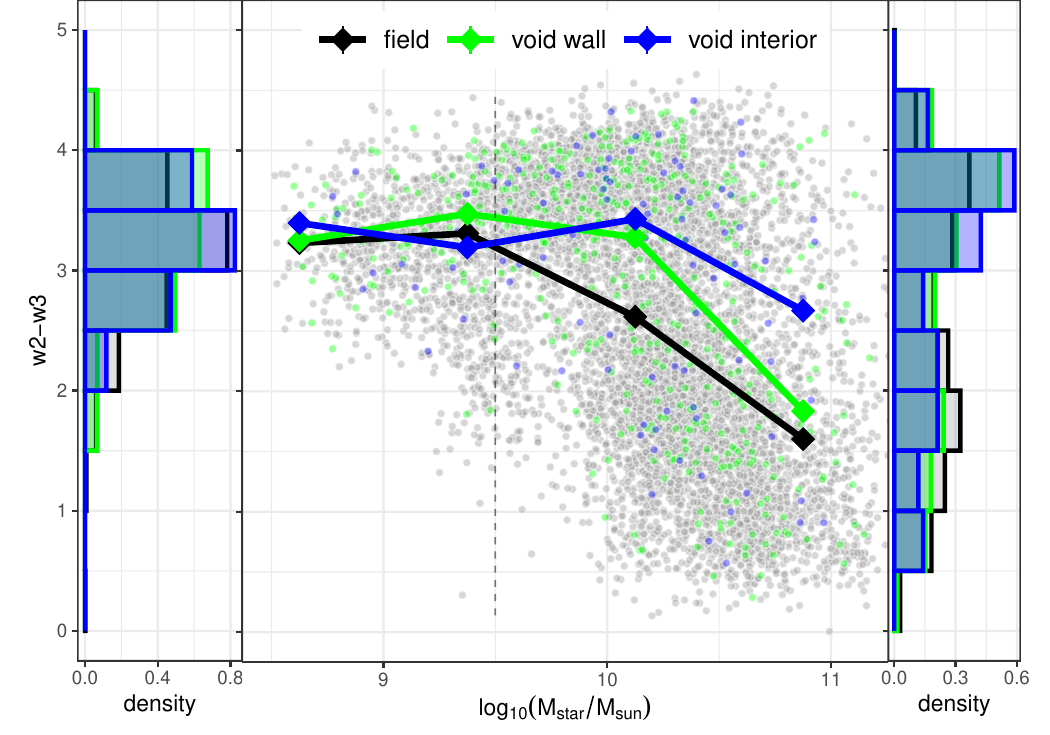}
    \caption{
    {\it{Central panel:}} Scatter plot of $w2-w3$ as a function of $log(M_*/M_{sun})$ for pair galaxies in 
voids (blue), in wall (green) and in fields (black).
Diamonds and solid lines indicate the median.  
The vertical dotted line represents $log(M_*/M_{sun}) =$ 9.5
{\it{Left panel:}} $g-r$ normalized distribution for $log(M_*/M_{sun}) <$ 9.5 paired galaxies in voids and field. 
{\it{Right panel:}} the same than left panel but for $log(M_*/M_{sun}) >$ 9.5. 
    }
    \label{fig:w2w3vsmstar}
  \end{subfigure}
  \hfill
  \begin{subfigure}[t]{0.49\textwidth}
    \includegraphics[width=\textwidth, height=0.75\textwidth]{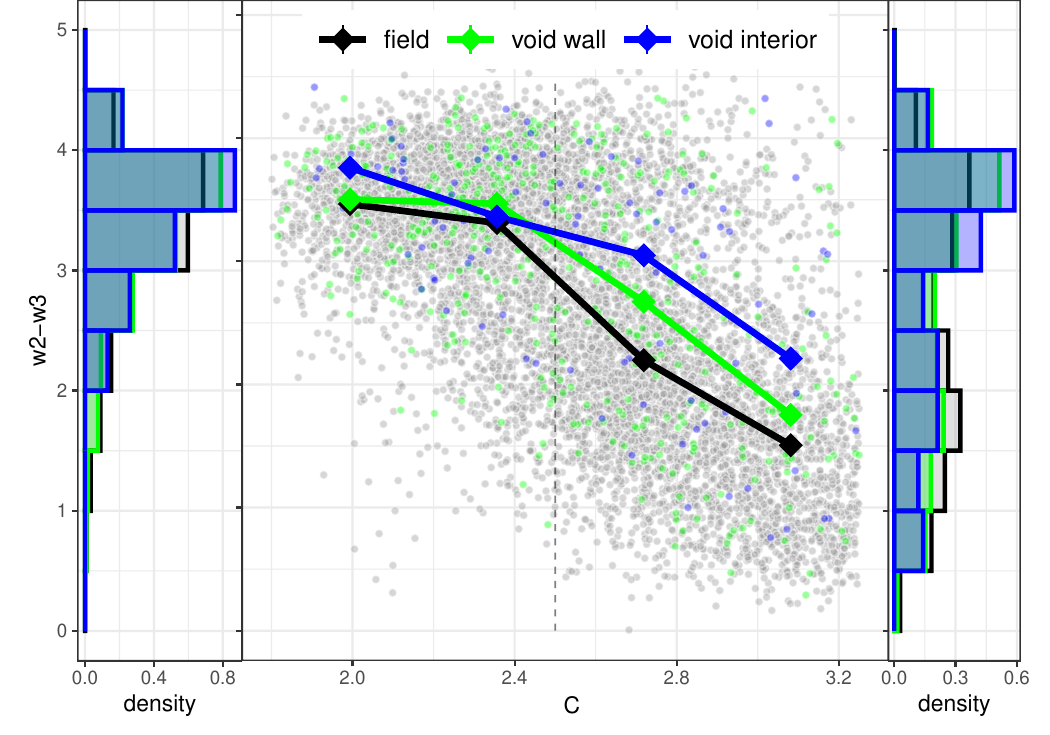}
    \caption{
    {\it{Central panel:}} Scatter plot of $w2-w3$ as a function of concentration index for 
pair galaxies in voids (blue), in wall (green) and in fields (black).
Diamonds and solid lines indicate the median.
{\it{Left panel:}} $g-r$ normalized distribution for $C <$ 2.5 paired galaxies in voids and field. 
{\it{Right panel:}} the same than left panel but for $C >$ 2.5. 
}
    \label{fig:w2w3vsCI}
  \end{subfigure}
   \caption{Study of stellar mass and concentration impact on IR color for void paired galaxies.  
   }
\end{figure*}

Since w2-w3 is linked to star formation activity, which is strongly affected by galaxy mass, we have analyzed the possible dependencies of these colors on stellar mass. In the middle panel of the Fig. \ref{fig:w2w3vsmstar} we show the color--stellar mass relation for pairs in void interiors (blue), void walls (green) and in the field (black), where the small points correspond to each galaxy in the matched samples and the diamonds and solid lines indicate the median value of $w2-w3$. The vertical dashed line indicates the value of stellar mass that divides the sample into low and high masses  ($log(M_*/M_{sun}) =$ 9.5).
It is notable that the colors are similar for low mass paired galaxies ($log(M_*/M_{sun}) < $ 9.5) in the analyzed global environments while they differ substantially for high mass paired galaxies ($log(M_*/M_{sun}) > $ 9.5), with void paired galaxies (blue and green lines) presenting larger colors than similar galaxies in the field (black lines).
In addition, low-mass galaxies present similar $w2-w3$ values, while for high-mass galaxies this tends to decrease if the mass increases.
In the left panel of the Fig. \ref{fig:w2w3vsmstar} we display the $w2-w3$ color distribution for low mass galaxies ($log(M_*/M_{sun}) < $ 9.5) and in the right panel the distribution corresponding to high mass galaxies ($log(M_*/M_{sun}) > $ 9.5).
In agreement with what we have shown in middle panel of this figure, the $w2-w3$ distributions are similar for low masses (comparable mean values and standard dispersions, also checked with the KS test) and differ for high masses.

Furthermore, the $w2-w3$ color is related to the morphology of the galaxies \citep{wright2010, jarrett17}, which leads us to examine the color variations with concentration. 
The results of the study of the color-concentration relation are displayed in Fig. \ref{fig:w2w3vsCI}, 
the analysis is similar to that carried out for the color--stellar mass relation (Fig. \ref{fig:w2w3vsmstar}).
In the middle panel of  Fig. \ref{fig:w2w3vsCI} we exhibit the $w2-w3$ color as a function of $C$ for the paired galaxies of the matched samples in void interior (blue), void walls (green) and in the field (red).
Similarly to the Fig. \ref{fig:w2w3vsmstar}, the diamonds represent the median values while the dots correspond to individual galaxies. The vertical black dashed line marks the threshold value of $C$ used to classify galaxies into low- and high-concentration categories ($C = 2.5$).
It can be seen in the figure that $w2-w3$ decreases as the concentration increases, this is observed for all the explored environments and is consistent with the tendency of the galaxies with higher concentrations to be less active star-formers than those with lower concentrations. 
We can also observed that low concentration paired galaxies ($C<2.5$) exhibit similar color-concentration relations in the large--scale environments analyzed, while high concentration galaxies ($C>2.5$) show notable differences with the environment, with void galaxies having higher colors than their field counterparts.
This result is reinforced by examining the $w2-w3$ color distributions for low-- and high--concentration paired galaxies in void interiors (blue), void walls (green), and in the field (red), displayed in the 
left and right panels of the Fig. \ref{fig:w2w3vsCI}, respectively.

\begin{table*} 
\center
\caption{Percentages of paired galaxies having w2-w3>2.5 selected accordingly mass and concentration, located in void interior environment, void walls and field. Standard errors are included.
}
\begin{tabular}{|c c c c c|}
\hline
Regions &  $log_{10}(M_{*}/M_{sun})<9.5 $ & $log_{10}(M_{*}/M_{sun})>9.5$ & $C < 2.5$ & $C > 2.5$  \\
\hline
\hline
Void interior &  90 $\pm$ 40 \% & 70 $\pm$ 20\% &  90 $\pm$ 30\% & 50 $\pm$ 20\%  \\
Void walls &  90 $\pm$ 20 \% & 60 $\pm$ 7\%  &  90 $\pm$ 10\% & 42 $\pm$ 7\%\\
Field & 87 $\pm$ 6 \% & 47 $\pm$ 1\%  &  86 $\pm$ 3\% & 30 $\pm$ 1\% \\
\hline
\hline
\end{tabular}
{\small}
\label{tab:wise}
\end{table*}

\subsection{Effects on stellar population and star formation activity}

\begin{figure}[ht]
\begin{center}
\includegraphics[width=0.43\textwidth]{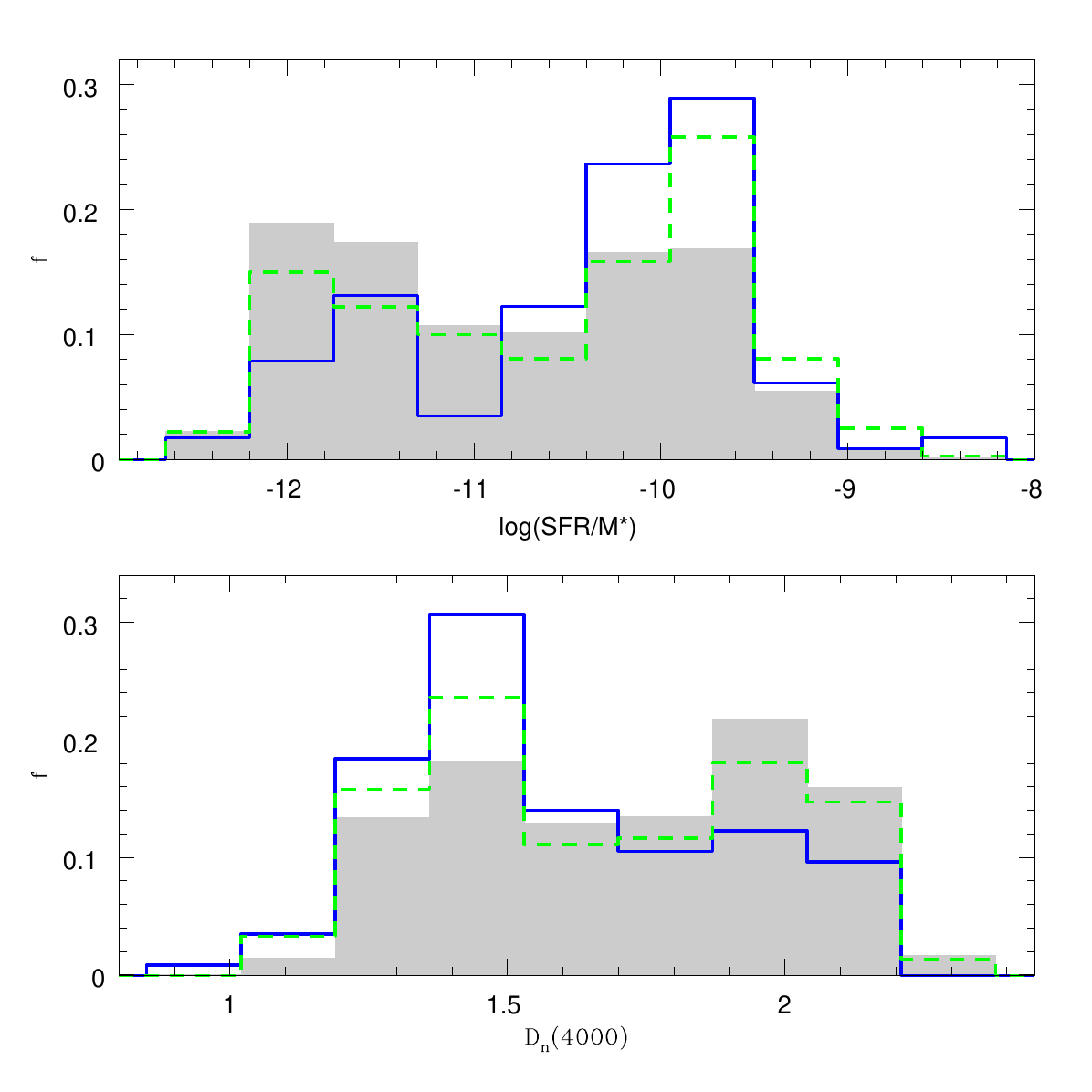}
\caption{Distributions of $log(SFR/M*)$ and $D_n(4000)$
for pair galaxies in void interiors (blue solid lines), in void walls (green dashed lines) and in fields (shaded histograms).
}
\label{HsfrDn}
\end{center}
\end{figure}

\begin{figure*}[ht]
  \begin{subfigure}[t]{0.49\textwidth}
    \includegraphics[width=\textwidth, height=0.75\textwidth]{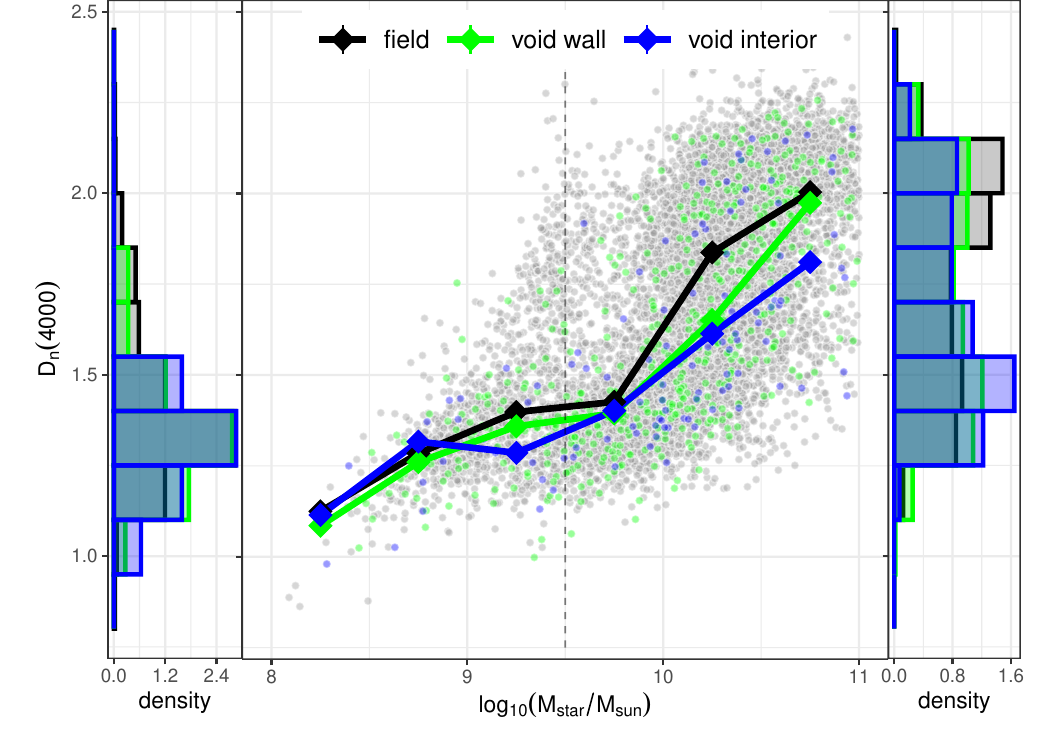}
    \caption{
    {\it{Central panel:}} Scatter plot of $D_n(4000)$ as a function of $log(M_*/M_{sun})$ 
    for pair galaxies in void interiors (blue), in void walls (green) and 
    in the field (black).
    Diamonds and solid lines indicate the median.
    The vertical dotted line represents $log(M_*/M_{sun}) =$ 9.5.
    {\it{Left panel:}} $D_n(4000)$ normalized distribution for $log(M_*/M_{sun}) <$ 9.5 paired galaxies 
    in voids and field. 
    {\it{Right panel:}} the same than left panel but for $log(M_*/M_{sun}) >$ 9.5. 
    }
    \label{fig:dn4000vsmstar}
  \end{subfigure}
  \hfill
  \begin{subfigure}[t]{0.49\textwidth}
    \includegraphics[width=\textwidth, height=0.75\textwidth]{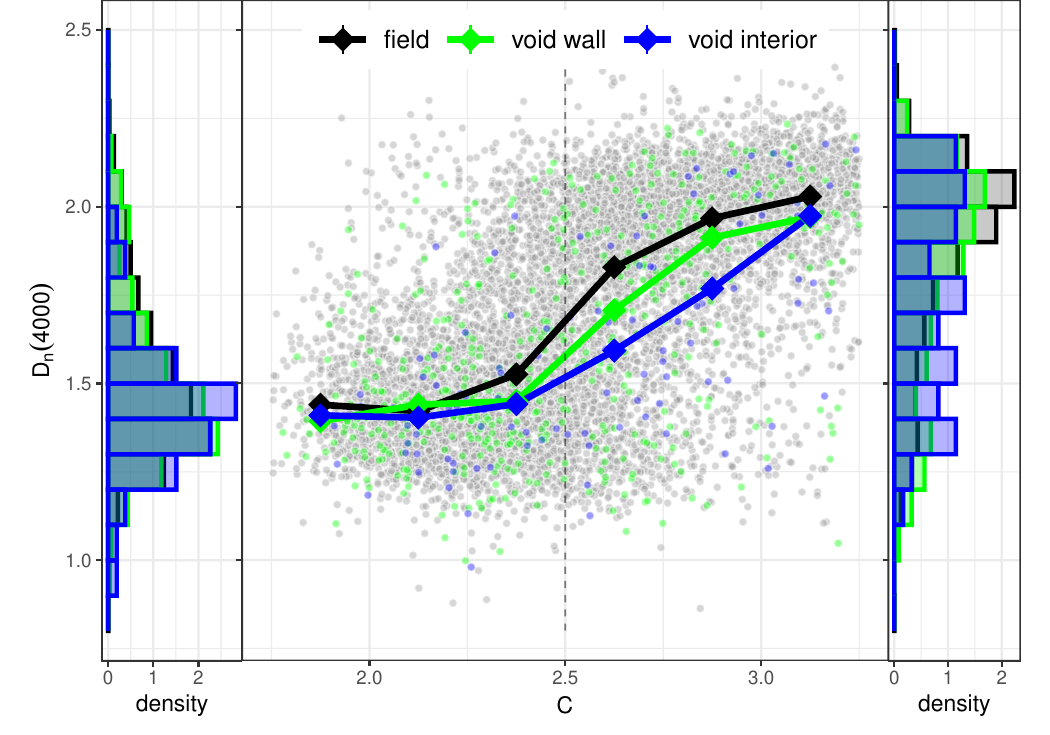}
    \caption{
    {\it{Central panel:}} Scatter plot of $D_n(4000)$ as a function of concentration index 
    for pair galaxies in void interiors (blue), in void walls (green) and 
    in the field (black).
    Diamonds and solid lines indicate the median.
    The vertical dotted line represents $C =$ 9.5.
    {\it{Left panel:}} $D_n(4000)$ normalized distribution for $C <$ 2.5 paired galaxies 
    in voids and field. 
    {\it{Right panel:}} the same than left panel but for $C >$ 2.5. 
    }
    \label{fig:dn4000vsc}
  \end{subfigure}
   \caption{Study of stellar mass and concentration impact on age of stellar population for void paired galaxies.  
   }
\end{figure*}

\begin{figure*}[ht]
  \begin{subfigure}[t]{0.49\textwidth}
    \includegraphics[width=\textwidth, height=0.75\textwidth]{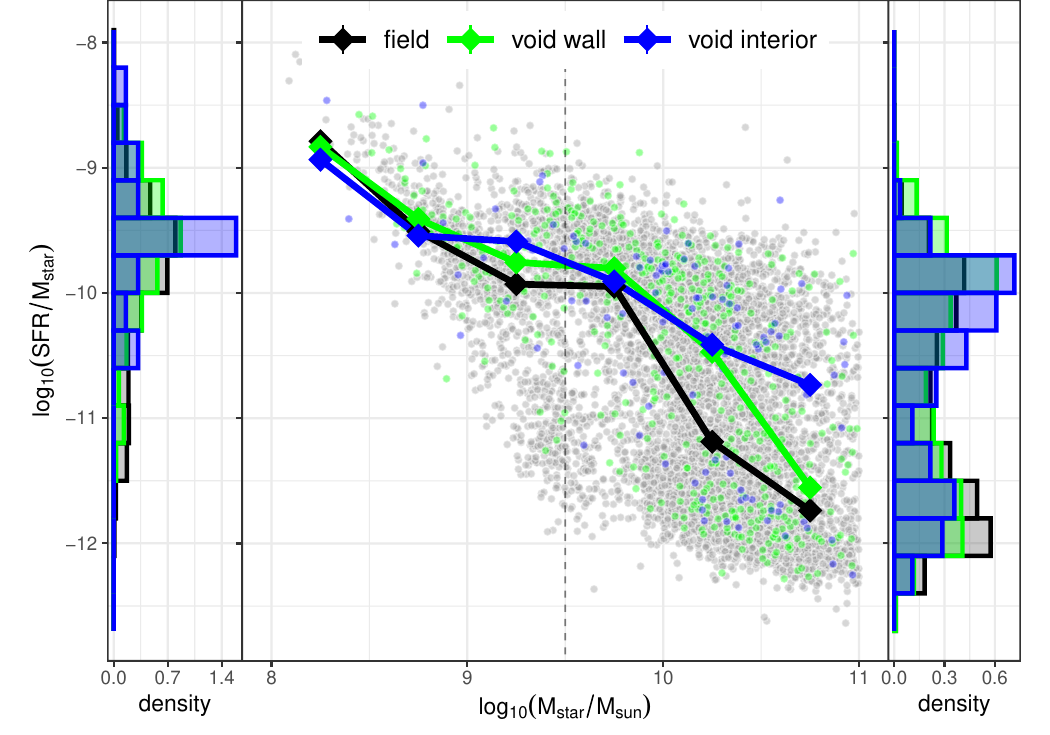}
    \caption{
    {\it{Central panel:}} Scatter plot of $log(SFR/M*)$ as a function of $log(M_*/M_{sun})$ 
    for pair galaxies in void interiors (blue), in void walls (green) and 
    in the field (black).
    Diamonds and solid lines indicate the median. 
    The vertical dotted line represents $log(M_*/M_{sun}) =$ 9.5.
    {\it{Left panel:}} $log(SFR/M*)$ normalized distribution for $log(M_*/M_{sun}) <$ 9.5 paired galaxies 
    in voids and field. 
    {\it{Right panel:}} the same than left panel but for $log(M_*/M_{sun}) >$ 9.5. 
   }
    \label{fig:sfrvsmstar}
  \end{subfigure}
  \hfill
  \begin{subfigure}[t]{0.49\textwidth}
    \includegraphics[width=\textwidth, height=0.75\textwidth]{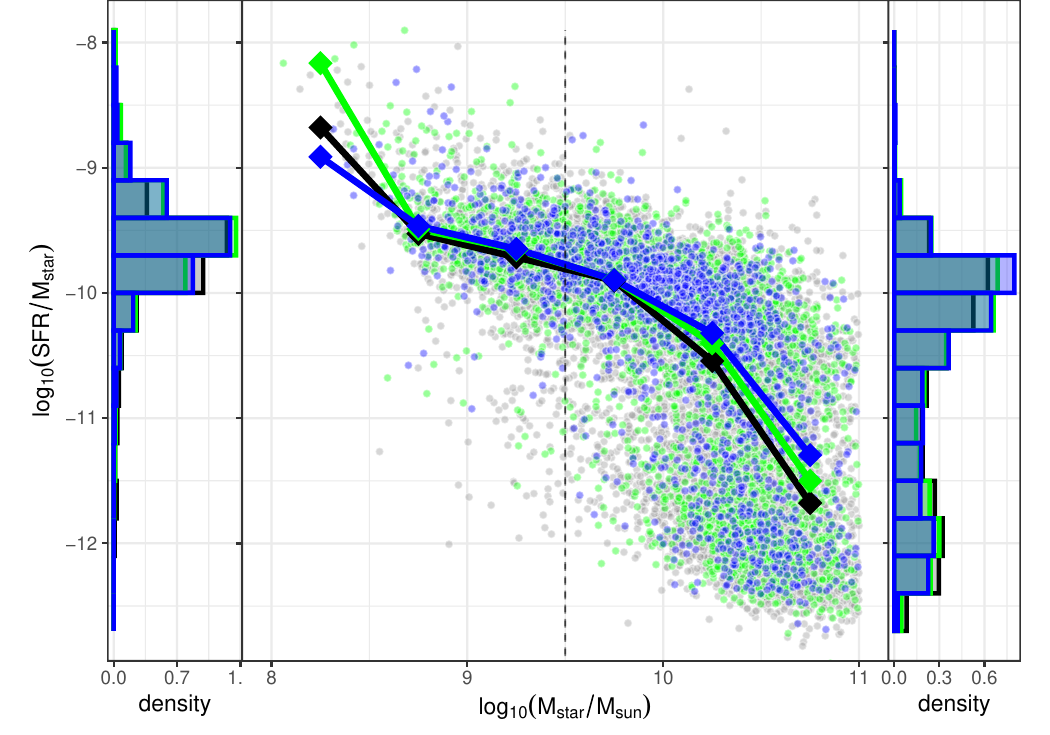}
    \caption{
     Analogous to panel (a) but for non-interacting galaxies
    }
    \label{fig:sfrvsmstari}
  \end{subfigure}
  \hfill
  \vspace{0.5cm} 
  \begin{subfigure}[t]{0.49\textwidth}
    \includegraphics[width=\textwidth, height=0.75\textwidth]{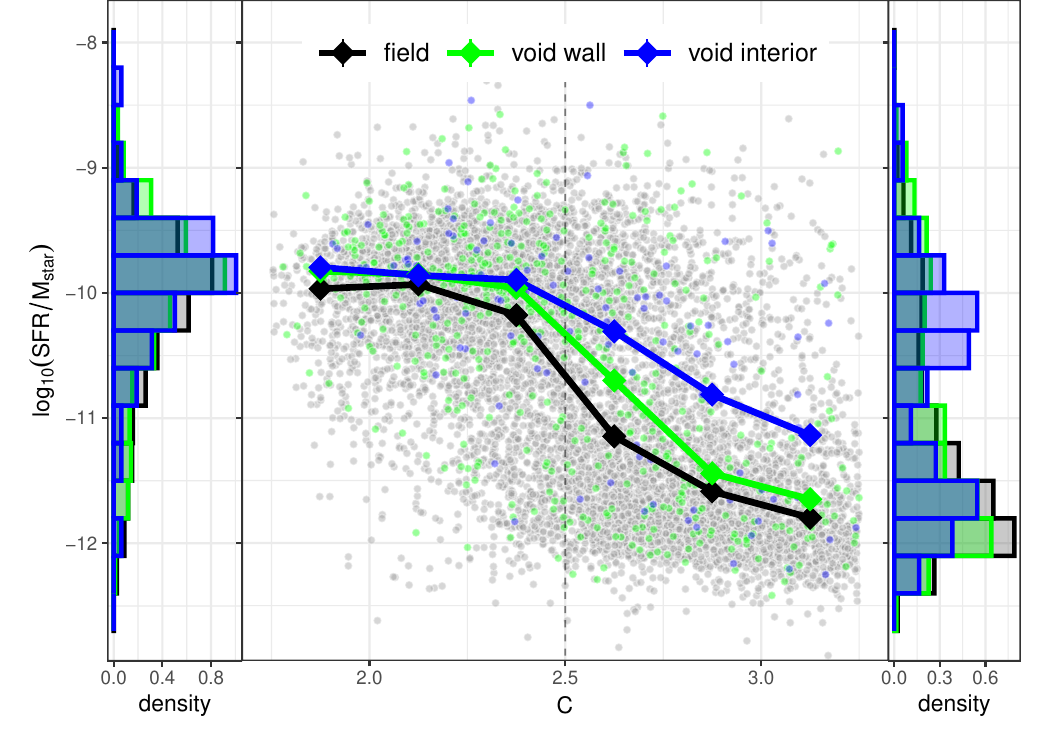}
    \caption{
    {\it{Central panel:}} Scatter plot of $log(SFR/M*)$ as a function of $C$ for pair galaxies in void interiors (blue), in void walls (green) and 
    in the field (black).
    Diamonds and solid lines indicate the median.
    The vertical dotted line represents $C =$ 9.5.
    {\it{Left panel:}}  $log(SFR/M*)$ normalized distribution for $C <$ 2.5 paired galaxies 
    in voids and field. 
    {\it{Right panel:}} the same than left panel but for $C >$ 2.5. 
    }
    \label{fig:sfrvsc}
    \end{subfigure}
   \hfill
  \begin{subfigure}[t]{0.49\textwidth}
    \includegraphics[width=\textwidth, height=0.75\textwidth]{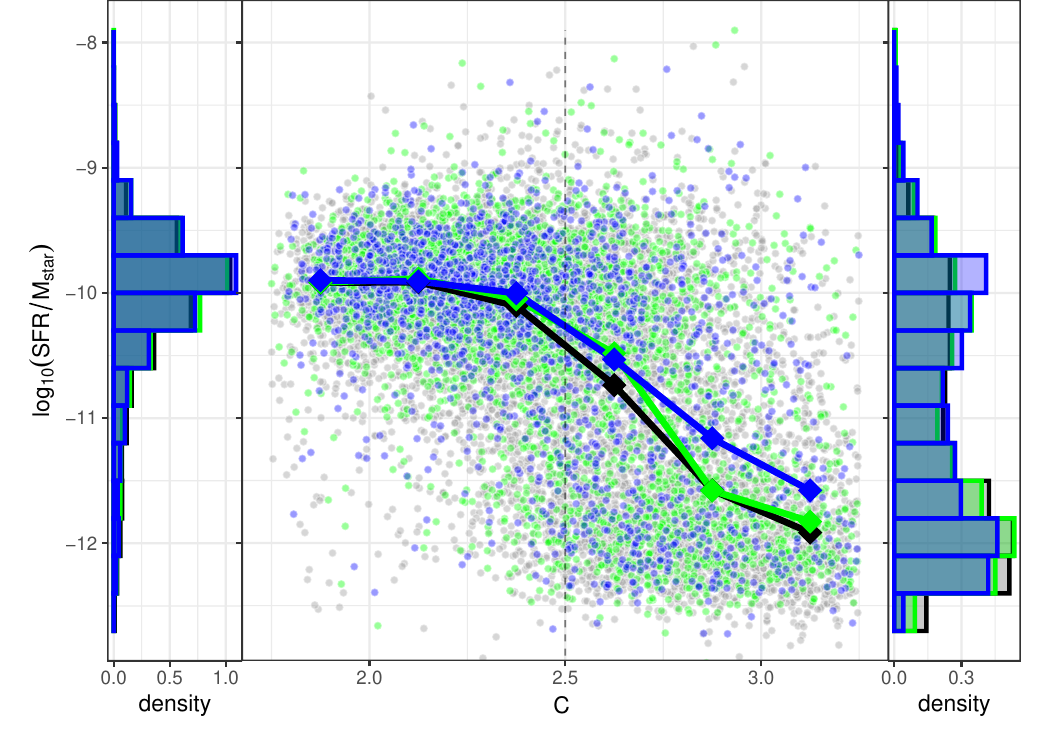}
    \caption{
    Analogous to panel (c) but for non-interacting galaxies
    }
    \label{fig:sfrvsci}
  \end{subfigure}
   \caption{Study of stellar mass and concentration impact on star formation activity for void paired galaxies. } 
   \label{fig:specsfrstudy}   
\end{figure*}

To evaluate the influence of the environment on the stellar age populations
of the paired galaxies residing in different  regions of cosmic voids, 
we employed the spectral index $D_{n}(4000)$  \citep{Kauffmann2003}, which assesses the age of stellar populations by calculating the spectral discontinuity at 4000 $\AA $. This discontinuity arises from the aggregation of numerous spectral lines within a narrow region of the spectrum, particularly significant in older stars. 
In our investigation, we adopted the $D_n(4000)$ definition outlined by \cite{Balogh1999}. This definition involves the ratio of average flux densities in narrow continuum bands (3850-3950 $\r{A}$ and 4000-4100 $\r{A}$), enabling a more precise determination of the ages of the stellar populations under study.

Additionally, we incorporated the specific star formation rate parameter, $log(SFR/M_*)$, 
which serves as a robust indicator of the star formation activity 
in the context of our analysis.
This parameter is computed as a function of the $H\alpha$ line luminosity and is normalized using stellar masses. The stellar masses considered in this study were determined through photometry fits by the MPA/JHU team, showing slight variations from those derived by \cite{Kauffmann2003} and \cite{Gallazi2005} using spectral indices, with the variances deemed negligible. Furthermore, the star formation rates ($SFR$) utilized in our analysis were extracted from the MPA/JHU catalogue, computed according to the methodologies outlined by \cite{Brinchmann2004}.

In order to provide an analysis of the influence of environmental density on galaxy interactions, we investigated the normalized distributions of $log(SFR/M_*)$ for paired galaxies situated in void, wall, and field regions (see upper panel in Fig. \ref{HsfrDn}).
It is evident that pair galaxies located in the interior regions of the cosmic voids exhibit higher levels of star formation activity in contrast to those located in the walls and fields.
In addition, pair systems within the field display lower values of $log(SFR/M_*)$, indicating comparatively less efficient star formation activity.
Furthermore, in a similar direction, the lower panel of Fig. \ref{HsfrDn} demonstrates that pairs within the interior regions of the voids present younger stellar populations (lower $D_n(4000)$ values) compared to their counterparts located in the wall and field. 
Nevertheless, pair galaxies situated in the field exhibit higher $D_n(4000)$ values, indicating a distinct trend toward older stellar populations.
The difference between these distributions was also quantified by the Kolmogorov-Smirnov statistics (with confidence of 99.8\%).
We can also observe that the value located near $log(SFR/M_*)$ $\approx$ -10.6 divides the distributions in two populations.
Similarly, galaxies in our samples show a bimodality in the stellar population around $D_n(4000)$ $\approx$ 1.6. 
Table \ref{tab:SFRDn} quantifies the percentages of galaxies with efficient star formation activity and a young stellar population in our samples, using these limits as reference. 
This finding suggests that pair galaxies located in voids exhibit higher star formation activity and younger stellar populations compared to pairs in the matched samples. 
The observation corroborates our previous findings, which highlighted an increased fraction of paired galaxies within cosmic voids exhibiting bluer colors, as discussed in preceding sections.

We have also explored the dependencies between the 
age of the stellar population of paired galaxies with their
stellar masses and concentration parameter.
Our results are shown in Figs. \ref{fig:dn4000vsmstar} and \ref{fig:dn4000vsc}, whose central panels display the relationship between $D_n(4000)$ with respect to $log(M_*/M_{sun})$ and $C$ parameter for pair galaxies located in the interior and in the wall of the cosmic voids (blue and green colors, respectively), and in the field regions (red colors); lines and diamonds indicate the $D_n(4000)$ median values whereas small points represent individual galaxies.  
In addition, we categorize samples into low and high--mass galaxies, using a threshold of $log(M_*/M_{sun})$=9.5 (vertical dashed black line in central panel of Fig. \ref{fig:dn4000vsmstar}). 
Similarly, we separate the samples into low and high concentration galaxies, depending on whether this is greater or less than $C=2.5$ (vertical dashed black line in central panel of Fig. \ref{fig:dn4000vsc}). 
As expected, a trend towards older stellar populations is evident as stellar mass and concentration increase for all the samples examined.
We observed that more massive galaxies with earlier morphological types (higher $C$ values) exhibit older stellar populations (central panels of the figures). 
We also note that galaxies in pairs, with low values of stellar mass and concentration, exhibit similar stellar age population, regardless of the galaxy pair position relative to their closest void.
On the other hand, paired galaxies with high stellar mass and concentration, show a remarkable disparity in the age of their stellar population, those galaxies in the inner regions of voids are characterized by young stellar populations, compared to pair systems positioned along the walls or in field region.
Additionally, the left and right panels of the Fig. 
\ref{fig:dn4000vsmstar} and \ref{fig:dn4000vsc}
present the normalized distributions of  $D_n(4000)$ for the different samples taking into account low and high stellar mass and concentration, which account for the observed behavior.

Moreover, we have explored the dependencies of star formation activity on mass and concentration and the results are shown in Fig. \ref{fig:sfrvsmstar} and \ref{fig:sfrvsc}, respectively. 
The central panels of the figures show 
the relation between the mean $log(SFR/M_*)$ 
as a function of stellar masses and concentration index of the pair galaxies in the void, wall and field regions.
As can be seen, star formation activity 
decreases toward higher stellar masses 
and earlier morphological types (higher $C$ values),
for all the samples studied in this work.

It is evident, from central panel of figure \ref{fig:sfrvsmstar},
that low-mass paired galaxies exhibit consistent star formation efficiency,
irrespective of their placement within the inner region of the cosmic voids, walls, or field.
Conversely, within the high-mass range, galaxies in paired systems situated within cosmic voids tend to display slightly higher levels of star formation activity 
in comparison to their counterparts located in the wall and in the field.
In addition, we categorize the samples into low and high-mass galaxies, using a threshold of $log(M_*/M_{sun}) = 9.5$. 
The figure \ref{fig:sfrvsmstar} illustrates the normalized distributions of
$log(SFR/M_*)$, 
encompassing both low and high mass ranges (left and right panels, respectively) for the different samples examined in this study, elucidating the observed trends.

As is it can be noticed from central panel of Fig. \ref{fig:sfrvsc}, 
early morphological types (high $C$ values) exhibit low star formation efficiency. 
We also note that galaxies in pairs with low values of $C$ parameter ($C<2.5$), indicating low concentration and corresponding to galaxies with spiral-type morphologies, exhibit similar levels of star formation activity, regardless of whether paired systems are situated within the inner regions of voids, on the walls or in the field environments.
Conversely, paired galaxies displaying high concentration, indicative of elliptical morphologies, demonstrate a notable contrast in their star-forming efficiency. Galaxies paired within the inner regions of voids display a distinct efficiency in star formation, compared to pair systems positioned along the walls or in field region.
Furthermore, we classify the samples into low and high-concentration galaxies, using a threshold of $ C = 2.5$. 
The left an right panels of figure \ref{fig:sfrvsc} presents the normalized distributions of $log(SFR/M_*)$ for the different samples taking into account low and high concentration parameter, which account for the observed behavior.
To quantify these findings, Table 5 presents the percentages of paired galaxies with efficient star formation and young stellar populations, selected accordingly stellar mass and concentration index, across void interiors, void walls, and the field. 

Additionally, we address the analysis of the effects of interactions on galaxies populating void regions.
We examine properties of pair galaxies compared to similar galaxies without close companions in similar large scale regions. 
Then, differences in the galaxy properties can be associated with pair interaction effects. 
This analysis allows us to examine the 
strength of interactions effects in underdense regions.

We selected galaxies in void interior, void walls and in the field, without neighbors at distances smaller 
than 150 h$^{-1}$ Mpc and radial velocity differences smaller than 500 km s$^{-1}$, hereafter non-interacting galaxies.
In order to perform appropriate comparisons we construct control samples of non--interacting galaxies having similar 
stellar lass, luminosity, redshift and concentration distributions than the corresponding pair galaxies.
The control samples of non--interacting galaxies encompass 1824, 3700 and 9029 objects in the void interior, void wall and field environments, respectively.
In figures \ref{fig:sfrvsmstari} and \ref{fig:sfrvsci} we display the dependencies of star formation activity 
on stellar mass and concentration for non-interaction galaxies. 
The figures are similar to those corresponding to paired galaxies. 

As it can be noticed from Figs. \ref{fig:sfrvsmstari} and \ref{fig:sfrvsci}, 
non-interacting galaxies in voids present a slight tendency to have more intense 
star formation activity than those that inhabit the field.
Nevertheless, the differences between void and field galaxies are more significant for those in pairs 
(Figs. \ref{fig:sfrvsmstar} and \ref{fig:sfrvsc})
compared to unpaired or non-interacting galaxies.
Galaxies in voids tend to exhibit stronger star formation activity than similar galaxies in the field, 
and this difference is especially enhanced when these galaxies are interacting with a nearby companion.

By selecting control samples with similar luminosities, redshift, concentration and stellar mass, in each large-scale region, the differences between paired and non-interacting galaxies can be attributed to their interactions.
Furthermore, for simplicity, we have only shown results for the star formation activity of non-interacting galaxies because these are qualitatively similar for the other astrophysical properties explored here.

\begin{table} 
\center
\caption{Percentages of paired galaxies with efficient star formation activity and young stellar populations, located in void environment and for matched samples of pairs in walls and fields. Standard errors are included.
}
\begin{tabular}{|c c c| }
\hline
Pair Restrictions &  $log(SFR/M_*)>-10.6$ &  $D_n(4000)<1.6$  \\
\hline
\hline
\% Void Interiors &  67.61 $\pm$ 0.79\% & 60.53 $\pm$ 0.73\%   \\
\% Void Walls &  57.52 $\pm$ 0.40\% & 50.82 $\pm$ 0.37\%  \\
\% Fields & 44.67 $\pm$ 0.67\% & 40.08 $\pm$ 0.63\%  \\
\hline
\hline
\end{tabular}
{\small}
\label{tab:SFRDn}
\end{table}

\begin{table*} 
\center
\caption{Percentages paired galaxies having efficient star formation and young stellar population, selected accordingly star mass and concentration, in void interior, void walls and field. Standard errors are included.
}
\begin{tabular}{|c c c c|}
\hline
Regions &  restrictions &  $log(SFR/M_*)>-10.6$ & $D_n(4000)<1.6$  \\
\hline
\hline
\% Void interior & $log_{10}(M_{*}/M_{sun})<9.5$ & 100. $\pm$ 20.\% & 100. $\pm$ 20.\%  \\
\% Void walls &    $log_{10}(M_{*}/M_{sun})<9.5$ & 93.  $\pm$ 9.\%  & 91. $\pm$ 9.\% \\
\% Field &         $log_{10}(M_{*}/M_{sun})<9.5$ & 82.  $\pm$ 2.\%  & 83. $\pm$ 2.\%\\
\hline
\% Void interior & $log_{10}(M_{*}/M_{sun})>9.5$ & 60. $\pm$ 6.\%  & 52. $\pm$ 5.\% \\
\% Void walls &    $log_{10}(M_{*}/M_{sun})>9.5$ & 50. $\pm$ 2.\%  & 44. $\pm$ 2.\% \\
\% Field &         $log_{10}(M_{*}/M_{sun})>9.5$ & 38.9 $\pm$ 0.8\% & 32.8 $\pm$ 0.7\% \\
\hline
\% Void interior & $C<2.5$ & 89. $\pm$ 12.\% & 89. $\pm$ 12.\%  \\
\% Void walls &    $C<2.5$ & 82. $\pm$ 5.\% & 75. $\pm$ 4.\% \\
\% Field &         $C<2.5$ & 76. $\pm$ 1.\%  & 70. $\pm$ 1.\%   \\
\hline
\% Void interior & $C>2.5$ & 49. $\pm$ 6.\% & 36. $\pm$ 4.\% \\
\% Void walls &    $C>2.5$ & 32. $\pm$ 2.\%  & 27. $\pm$ 1.\% \\
\% Field &         $C>2.5$ & 21.6 $\pm$ 0.6\% & 17.1 $\pm$ 0.5\% \\
\hline
\hline
\end{tabular}
{\small}
\label{tab:spec}
\end{table*}

\subsection{Contribution of pairs to the star formation efficiency in voids }

Finally, we have estimated the contribution of paired galaxies to star formation in voids.
We have selected volume--limited samples of paired and non--interacting galaxies in voids, walls, and in the field. These samples are defined by a maximum redshift (z = 0.1) and a limiting luminosity in the r band ($M_r$ = -19.6).
We have considered different large--scale regions depending on the void centric distance,
as detailed below,\\

- $inner-void$: from 0 to 0.6 $R_v$

- $outer-void$: from 0.6 to 0.8 $R_v$

- $void-wall$: from 0.8 to 1.2 $R_v$

- $field$: from 1.8 to 2.5 $R_v$.\\

\begin{figure}[ht]
\begin{center}
\includegraphics[width=0.45\textwidth]{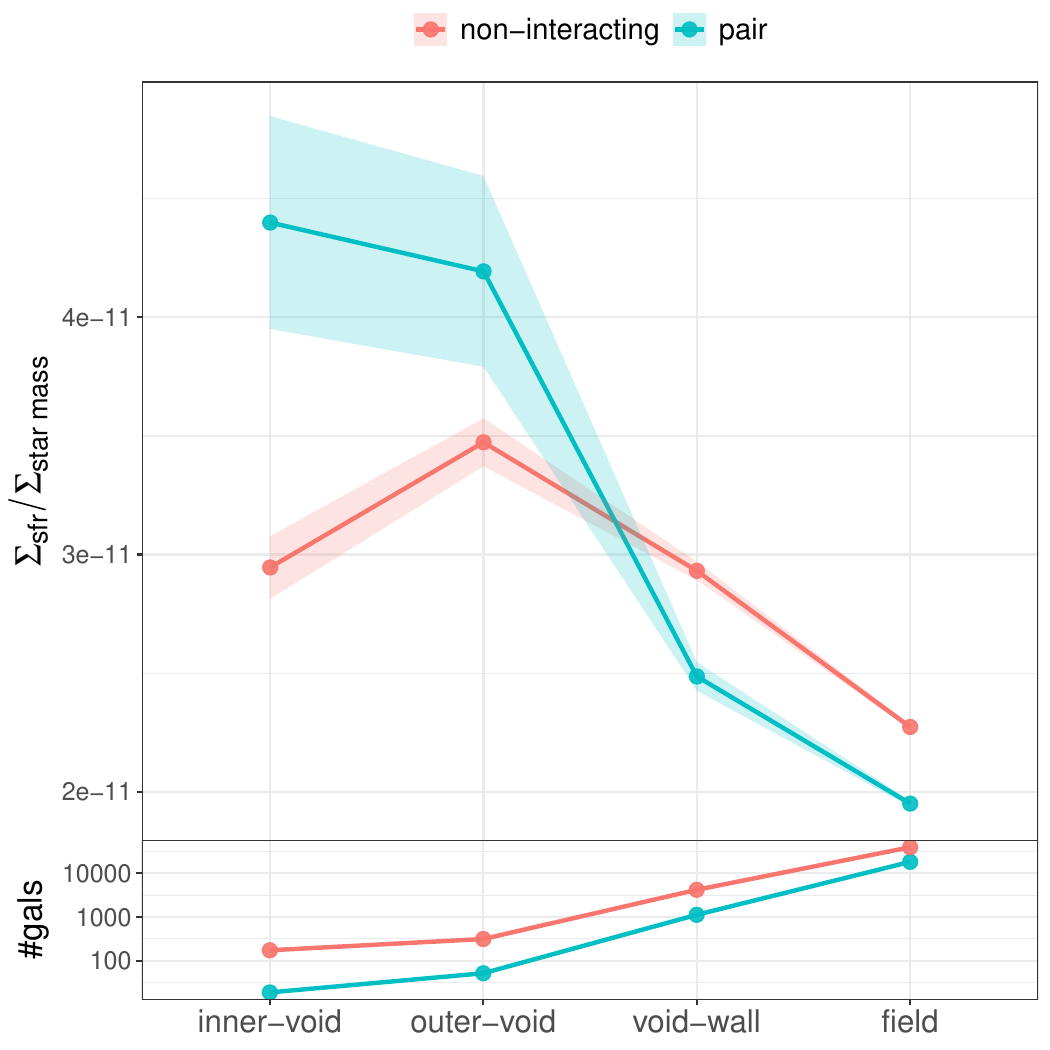}
\caption{{\it{Upper panel:}} star formation efficiency in void regions and field for paired and non-interacting galaxies. Shaded regions correspond to standard errors.
{\it{Lower panel:}} number of galaxies in the volume--complete samples.
}
\label{effi}
\end{center}
\end{figure}

The numbers of galaxies corresponding to each sample are represented in lower panel of Fig. \ref{effi}, red color corresponds to paired galaxies and  light blue symbolizes non--interacting galaxies.

To quantify the relative contribution of galaxy pairs to star formation we introduce the efficiency, which is calculate
by adding the stellar masses formed by each galaxy in the sample divided by the total stellar mass of these galaxies, 

  \begin{equation}
   e=\frac{\sum SFR}{\sum M_{star}/M_{sun}}
   \label{eq:e}
   \end{equation}

The results are portrayed in upper panel of the Fig. \ref{effi}, where red line corresponds to pairs and  light blue line indicates non--interacting galaxies. As it can be seen, in void interior regions (inner-- and outer-void) star formation is significantly more efficient in paired galaxies than in non--interacting ones. 
On the other hand, in regions where global density increases, non-interacting galaxies show a higher efficiency in their star formation activity than pairs.


\section{Discussion and conclusions}

In this work we have analysed astrophysical properties of galaxy pairs in cosmic voids.
The particular environment of voids, in both galaxy/gas density and dynamics differ substantially from elsewhere providing suitable conditions to study the pure impact of galaxy interactions with weaker global perturbations.
In fact, objects inside cosmic voids have preferentially expanding radial trajectories with small transversal motions. Thus, pair members are expected to have significantly different orbital behavior and evolution which would reflect in present-day astrophysical properties. 
In our study, we focus on the comparison between galaxy pairs in voids, wall interphase, and the global mean environment, field.
For this aim, we construct matched pair samples that allow to study galaxy properties such as optical colors, mid infrared photometry, and spectroscopically derived star formation rates and stellar population ages.
The matched wall and field pair samples are constrained to fit the observed void galaxy pairs mean redshift, r--band luminosity, stellar mass and light concentrations parameter.
We find that optical colors show pair members in voids to be significantly bluer than the corresponding pair members residing in wall an field. This difference does not depend strongly on galaxy stellar mass nor concentration nor r--band luminosity.
WISE mid infrared photometry for a subsample of pairs in the three environments offers the opportunity of studying the  remission by the dusty environment associated to star forming regions in galaxies. We obtain significantly larger w1-w2 and w2-w3 values for the void pair members with respect to the corresponding matched samples in the wall and in field environments. By contrast with optical colors, mid IR  photometry shows a dependence with both stellar mass and concentration parameter. The larger differences are observed for high mass and concentration.
We also notice that mid IR color-color diagram shows void pair members consistent with star forming galaxies, in contrast with the other environments that exhibit a bimodal behavior of passive and star forming objects.
This fact reflects a large star formation activity in pairs residing in cosmic voids.
The stellar population age parameter, D$_n$(4000), shows a significant younger population associated to pair members in voids in line with the findings for the mid IR colors. This is also reflected in the higher star formation efficiency $SFR/M_*$ values which show a larger efficiency for pairs residing in voids.
We also notice that the star formation efficiency is larger for luminous pair members in cosmic voids.
The lack of a significant star formation activity enhancement in the faint members may originate in the fact that these galaxies are intrinsically more unstable to generate new stars which makes them more difficult to distinguish the different environments.
A comparison between pair members and non-interacting galaxies show a soft tendency to have bluer optical colors and larger star formation rates, but this weak effect is not so clearly evident as in the comparison between the pair populations.
Globally speaking, we also find that the efficiency of star formation associate to galaxy interactions is significantly larger in cosmic void pairs. This larger star formation activity could be associated to both the richer gas environment and the expected more gentle behavior as compared to the more eccentric orbital dynamics and strong interactions and mergers more likely to have taken place in wall and field environments.

\begin{acknowledgements}
This work has been partially supported by Consejo de Investigaciones 
Cient\'{\i}ficas y T\'ecnicas de la Rep\'ublica Argentina (CONICET), the
Secretar\'{\i}a de Ciencia y T\'ecnica de la Universidad Nacional de C\'ordoba (SeCyT) and Secretar\'ia de Ciencia y T\'ecnica de la Universidad Nacional de San Juan.             
Funding for the SDSS and SDSS-II has been provided by the Alfred P. Sloan Foundation, the Participating Institutions, the National Science Foundation, the U.S. Department of Energy, the National Aeronautics and Space Administration, the Japanese Monbukagakusho, the Max Planck Society, and the Higher Education Funding Council for England. The SDSS Web Site is http://www.sdss.org/. The SDSS is managed by the Astrophysical Research Consortium for the Participating Institutions. The Participating Institutions are the American Museum of Natural History, Astrophysical Institute Potsdam, University of Basel, University of Cambridge, Case Western Reserve University, University of Chicago, Drexel University, Fermilab, the Institute for Advanced Study, the Japan Participation Group, Johns Hopkins University, the Joint Institute for Nuclear Astrophysics, the Kavli Institute for Particle Astrophysics and Cosmology, the Korean Scientist Group, the Chinese Academy of Sciences (LAMOST), Los Alamos National Laboratory, the Max-Planck-Institute for Astronomy (MPIA), the Max-Planck-Institute for Astrophysics (MPA), New Mexico State University, Ohio State University, University of Pittsburgh, University of Portsmouth, Princeton University, the United States Naval Observatory, and the University of Washington.
This research has used the NASA's Astrophysics Data System. 
Plots were performed using R software.

\end{acknowledgements}

\section*{DATA AVAILABILITY}
The data underlying this article will be available upon reasonable request to the corresponding author.

%
%

\bibliographystyle{aa} 
\bibliography{biblio.bib} 

\end{document}